\documentclass[aps,prx,superscriptaddress,twocolumn]{revtex4-2}
\usepackage{graphicx,amsmath,amssymb,amsfonts,latexsym,color,dcolumn,bm}

\newcommand{\beq}{\begin{equation}}
\newcommand{\eeq}{\end{equation}}
\newcommand{\bea}{\begin{eqnarray}}
\newcommand{\eea}{\end{eqnarray}}

\usepackage{amsfonts,amssymb}
\usepackage{dsfont}
\usepackage{physics}
\usepackage{tocvsec2}

\usepackage{appendix}

\usepackage{tikz}

\usepackage{amsmath}
\usepackage{bbold}
\usepackage{hyperref}
\hypersetup{
     colorlinks=true,      
    linkcolor=blue,        
    citecolor=blue,        
    filecolor=magenta, 
    urlcolor=black          
}

\usepackage{comment}

\begin{document}

\title{Spectral single photons characterization using generalized Hong-Ou-Mandel interferometry}

\author{N. Fabre\footnote{nfabre@ucm.es}}
\affiliation{Centre for Quantum Optical Technologies, Centre of New Technologies, University of Warsaw, ul.  Banacha 2c, 02-097 Warszawa, Poland}
\affiliation{Departamento de Óptica, Facultad de Física, Universidad Complutense, 28040 Madrid, Spain}

\date{\today}
\begin{abstract}
We propose new methods to characterize the spectral-temporal distribution of single photons. The presented protocols take advantage of spectral filtering, frequency entanglement between two single photons, the one of interest and a reference, followed by the generalized Hong-Ou-Mandel interferometer. The measurement of the coincidence probability in these different schemes reveals the chronocyclic Wigner distribution, the pseudo-Wigner one and the spectrogram at the single photon level.
 \end{abstract}
\vskip2pc 

\maketitle

\section{Introduction}
Quantum information can be encoded by using the degrees of freedom of single photon fields. Such degrees of freedom are described by discrete variables corresponding to the polarization, spatial path, orbital angular momentum or by continuous variables, such as transversal position-momentum \cite{tasca_continuous_2011} and time-frequency \cite{fabre_generation_2020}. Discrete variables encoding of single photons have been used in many quantum protocols and the most recent achievements are the experimental realization of boson sampling \cite{spring_boson_2013,noauthor_photonic_2019} and in quantum communication protocols \cite{yin_entanglement-based_2020}. We point out that time-frequency continuous variables degree of freedom of single photons have, for now, been marginally used in experimental quantum protocols. We can find  in the literature instances as a quantum key distribution protocol \cite{zhong_photon-efficient_2015}, and recently in Hong, Ou and Mandel (HOM) metrology \cite{chen_hong-ou-mandel_2019,lyons_attosecond-resolution_2018}. Indeed, such infinite dimensional Hilbert spaces are often discretized into modes \cite{kues_-chip_2017,imany_high-dimensional_2019,maltese_generation_2020,raymer_temporal_2020,PhysRevX.5.041017} relevant for discrete variables quantum protocols. Besides, the manipulation of the frequency degree of freedom requires non-linear interaction at the single photon level which is an experimental challenge. However, the recent development of optical schemes that can manipulate the frequency of single photons has attracted significant interest for this degree of freedom. Indeed, it is a robust one because there is no linear physical mechanism that affects the spectral distribution of single photons. Thus, frequency encoding combined with integrated optical technologies (see for instance \cite{appas_flexible_2021,luo_counter-propagating_2020,joshi_frequency-domain_2020}), is now considered to be relevant, especially because it offers a drastic reduction in resources requirements for all-photonic quantum networks.\\

Characterizing or achieving the spectral tomography of single photons is a difficult task. Optical techniques  measuring the amplitude and phase of the spectral distribution of classical fields are difficult to apply directly to single photons because they require non-linear interactions. The frequency-resolved optical grating (FROG) \cite{Kane:93}, and spectral phase interferometry for direct electro-field reconstruction (SPIDER)  \cite{Wong:94} are examples of these techniques. Hence, new proposals have been proposed to measure the spectral distribution of single photons. For instance, the coincidence probability obtained with the generalized HOM interferometry allows the measurement of the phase-matching function of a photon pair produced by spontaneous parametric down-conversion (SPDC) \cite{douce_direct_2013,boucher_toolbox_2015}. Other techniques have been investigated to measure the spectral properties of photon pairs such as shear interferometry \cite{davis_measuring_2020,gianani_robust_2019,doi:10.1116/1.5136340}, the measurement of four marginals of the chronocyclic Wigner distribution of biphoton states in \cite{PhysRevA.100.033834} or an intensity interferometry scheme with a chirp pulse \cite{thekkadath_measuring_2022}. Spectral-temporal tomography of single photons has also been realized, with a compressive technique measuring the chronocyclic Wigner distribution through its mode decomposition \cite{gil-lopez_universal_2021}, direct measurement of the density matrix with a spectrally-resolved HOM interferometer \cite{thiel_single-photon_2020} and by interfering with a known reference field and the single photon of interest \cite{huisman_instant_2009,zavatta_tomographic_2004,wasilewski_spectral_2007}. \\

In this paper, we propose a new method to perform the tomography of the spectral-temporal distribution of single photons. The idea consists of generating frequency entanglement between the single photon state of interest and a reference one. Then, the entangled state enters into the generalized HOM interferometer. The measurement of the coincidence probability is the chronocyclic Wigner distribution \cite{brecht_characterizing_2013,fabre_generation_2020}, which thus corresponds to the full tomography of the state. If spectral filtering operation is performed before the single photon state of interest gets frequency entangled, the chronocyclic pseudo-Wigner distribution \cite{shin_pseudo_1993,cohen_1989} is obtained using this method. In the two cases, the reference state intervenes during the entanglement gate and also to assists in the HOM interference effect. Its spectral structure is not relevant and does not have to be known. In addition, we analyze the expression of the coincidence measurement of the generalized HOM interferometer when two separable single photons are impinged into the beam-splitter. The coincidence probability corresponds to the spectrogram \cite{dorrer_concepts_2005} only when the state is pure. Our approach relies directly on non-linear interactions at the single photon level, which is legitimized by the development of new quantum materials and optical schemes that manipulate the frequency degree of freedom. In addition, the frequency entanglement gate introduced in this paper is expressed formally in terms of time and frequency operators which were introduced in \cite{fabre:tel-03191301}, and more rigorously in \cite{maccone_quantum_2020}. This formulation emphasizes the mathematical analogy between the quadrature position-momentum formalism and the time-frequency degree of freedom of single photons. This mathematical approach allows laying the path toward time-frequency continuous variables encoding relying on non-linearity to entangle deterministically single photons.   \\

The paper is organized as follows. In Sec.~\ref{sectiontwo} , we consider the description of a single photon state with a Hilbert and phase space formalism. In Sec.~\ref{sectionthree}, we introduce a method for measuring the chronocyclic Wigner distribution Sec.~\ref{measurementFBS}, the chronocyclic pseudo-Wigner distribution Sec.~\ref{pseudonext} and Sec.~\ref{othertech}, and the spectrogram Sec.~\ref{spectrogramsec} at the single photon level using generalized HOM interferometry.  Finally, we recap the results in Sec.~\ref{conclusion} and provide additional perspectives.

\section{Single photon formalism and chronocyclic phase space distribution}\label{sectiontwo}
In this section, we describe a single photon state with a spectral distribution. We will choose to work with the frequency variable but note that at the single photon level, such a variable is related to the energy by the relation $E=\hbar \omega$. \\

A single photon with frequency $\omega$ and in spatial mode $a$ is described by the creation operator applied to the vacuum state $|0\rangle$ as follows: $\hat{a}^{\dagger}(\omega)|0\rangle=|\omega\rangle$. The annihilation operator at frequency $\omega$ is defined as $\hat{a}(\omega)|\omega'\rangle=\delta(\omega-\omega')|0\rangle$. The canonical conjugate continuous variable of the frequency is the time of arrival variable and is defined by the Fourier transform: $\hat{a}^{\dagger}(t)=\int_{\mathbb{R}} e^{i\omega t} \hat{a}^{\dagger}(\omega) d\omega$. A single photon with time of arrival $t$ is then: $\hat{a}^{\dagger}(t)|0\rangle=|t\rangle$. Since $|\omega\rangle$ and $|t\rangle$ are orthogonal basis, a single photon wave function can be decomposed as
\begin{equation}\label{wavefunctionsingle}
|\psi\rangle=\int_{\mathbb{R}} d\omega \psi(\omega) |\omega\rangle =\int_{\mathbb{R}} \tilde{\psi}(t)dt |t\rangle,
\end{equation}
where $\psi(\omega)$ is the amplitude spectrum of the source and its Fourier transform $\tilde{\psi}(t)$ is the amplitude of the time of arrival distribution. The normalization of the wave function gives the relations: $\int_{\mathbb{R}} d\omega |\psi(\omega)|^{2}=\int_{\mathbb{R}} dt |\tilde{\psi}(t)|^{2}=1$. The integration range is extended to $\mathbb{R}$ in the two domains. For the frequency variable, one can argue that the spectral distribution of interest is sufficiently peaked around the THz frequency for the considered quantum optics applications  and far away from the zero and negative frequencies. The general density matrix of a spectral single photon state can be written as
 \begin{equation}\label{densitymatrix}
 \hat{\rho}=\iint_{\mathbb{R}} d\omega d\omega' \rho(\omega,\omega') |\omega\rangle\langle\omega'|.
 \end{equation}
 where the function $\rho$ verifies $\iint_{\mathbb{R}} d\omega d\omega' \rho(\omega,\omega') =1$ from $\text{Tr}(\hat{\rho})=1$. The pure case is recovered when the function $\rho$ is separable, namely, $\rho(\omega,\omega')=\psi(\omega)\psi^{*}(\omega')$.
We introduce time and frequency operators \cite{fabre:tel-03191301} defined as
\begin{align}
\hat{t}_{a}=\int_{\mathbb{R}} dt\ t \ \hat{a}^{\dagger}(t)\hat{a}(t),\\
\hat{\omega}_{a}=\int_{\mathbb{R}} d\omega \ \omega \ \hat{a}^{\dagger}(\omega)\hat{a}(\omega).
\end{align}
These operators admit as eigenvalues the time of arrival and the frequency of a single photon state: $\hat{t}|t\rangle=t|t\rangle$ and $\hat{\omega}|\omega\rangle=\omega|\omega\rangle$. Such operators do not commute: $[\hat{t},\hat{\omega}]=i\mathbb{I}$, and thus they are mathematically fully analogous to the position-momentum operators as they obey the Heisenberg algebra. Such a mathematical analogy will be useful for writing frequency entanglement single photon gates in the next section Sec.~\ref{sectionthree}. Such time and frequency operators will be relevant for the reasons explained later. A rigorous form of time and frequency operators was introduced in \cite{giovannetti_quantum_2015,maccone_quantum_2020,fabre:tel-03191301}  and involves a quantum clock, emphasizing that the measurement of a single photon of time of arrival $t$ is conditioned to the clock has clicked from a single photon event. Such a formalism is not needed in this study.

Instead of using a Hilbert space description, a single-photon state can be described with a phase space distribution called the chronocyclic Wigner (CW) one: \cite{brecht_characterizing_2013,fabre_generation_2020}
\begin{equation}
W_{\hat{\rho}}(\omega,t)=\frac{1}{\pi} \int_{\mathbb{R}} d\omega' e^{2i\omega' t}\langle\omega-\omega'|\hat{\rho}|\omega+\omega'\rangle,
\end{equation}
 which is normalized to one as a consequence of the normalization of the density matrix. 
Using Eq.~(\ref{densitymatrix}), we can write the distribution as
 \begin{equation}
 W_{\hat{\rho}}(\omega,t)=\frac{1}{\pi} \int_{\mathbb{R}} d\omega' e^{2i\omega' t} \rho(\omega-\omega',\omega+\omega').
 \end{equation}
If the state is pure: $\hat{\rho}=|\psi\rangle\langle\psi|$ (see Eq.~(\ref{wavefunctionsingle})) we obtain
 \begin{equation}
W_{\psi}(\omega,t)=\frac{1}{\pi} \int_{\mathbb{R}} d\omega' e^{2i\omega' t} \psi(\omega-\omega')\psi^{*}(\omega+\omega').
\end{equation}
The chronocyclic Wigner distribution  is quadratic in the spectral distribution $\psi$ and contains information about both the amplitude and phase of $\psi$. Such a distribution can possess negative values;  thus, it is a quasi-probability distribution. The marginals of the chronocyclic Wigner distribution are positive and experimentally measurable quantities: 
\begin{align}
\int_{\mathbb{R}} d\omega W_{\psi}(\omega,t)= |\tilde{\psi}(t)|^{2},\label{timemarg}\\
\int_{\mathbb{R}} dt W_{\psi}(\omega,t)= |\psi(\omega)|^{2}.
\end{align}
The time-of-arrival distribution defined in Eq.~(\ref{timemarg}) can be measured with a time-resolved single photon detector \cite{luo_counter-propagating_2020}, but this measurement can be challenging for single photons possessing a temporal bandwidth lower than the picosecond range, produced by certains integrated optical waveguides \cite{francesconi_engineering_2020,maltese_generation_2020}. Nevertheless, the intensity frequency distribution can be measured using an optical fiber that maps the time of arrival to the frequency variable. Full tomography can be performed by measuring all rotated marginals, meaning by performing a fractional Fourier transform \cite{fabre:tel-03191301}, analogously to the reconstruction procedure of the quadrature variable state with the homodyne detection. However, this technique also requires detectors with a high temporal resolution. Finally, the reconstruction of the spectral wave function from the chronocyclic Wigner distribution is:
\begin{equation}\label{reconstruction}
\psi(\omega)=\frac{1}{\psi^{*}(0)} \int_{\mathbb{R}} W_{\psi} (\frac{\omega}{2},t)e^{i\omega t} dt.
\end{equation}
We now introduce the chronocyclic pseudo-Wigner distribution (CPW) \cite{allard_utilisation_1987} written here in the quantum formalism:
\begin{equation}\label{PseudoWigner}
PW_{\hat{\rho}}(\tau,\mu)=\int_{\mathbb{R}} d\omega f(-\omega)f^{*}(\omega) \langle\mu-\omega|\hat{\rho} |\mu+\omega\rangle e^{-2i\omega\tau}.
\end{equation}
By considering the window function $f$ even and real, and considering again the pure case $\hat{\rho}=|\psi\rangle\langle\psi|$, the Eq.~(\ref{PseudoWigner}) is reduced to
\begin{equation}
PW_{f\psi}(\tau,\mu)=\int_{\mathbb{R}} d\omega f^{2}(\omega)\psi(\mu-\omega)\psi^{*}(\mu+\omega)e^{-2i\omega\tau}.
\end{equation}
The CPW distribution is related to the chronocyclic Wigner distribution by $PW_{f^{2}\psi}(\tau,\mu)=W_{\psi_{t}}(\tau,\mu)$, where $\psi_{t}(u)=\psi(u)f(u-t)$ \cite{Wignerville}. In other words, the CPW distribution is the convolution of the chronocyclic Wigner distribution of the single photon and the one of the reference one used as a window:
\begin{equation}
PW_{f\psi}(\tau,\mu)=\frac{1}{\pi}\int_{\mathbb{R}} d\omega W_{\psi}(\tau,\omega)W_{f}(0,\mu-\omega).
\end{equation}
Then, the chronocyclic Wigner distribution can be deduced from the relation:
\begin{equation}\label{recoverWigner}
W_{\psi}(\tau,\omega)= \int_{\mathbb{R}} e^{i\omega t} \frac{\tilde{PW}_{f\psi}(\tau,t)}{\tilde{W}_{f}(0,t)} dt,
\end{equation}
where the tilde notation $\tilde{PW}_{f\psi}(\tau,t),\tilde{W}_{f}(0,t)$ corresponds to the Fourier transform of the CPW and CW distribution respectively. For computational reasons, the CPW distribution has been popularly used instead of the CW distribution for the time-frequency analysis of classical signals. The reconstruction of an unknown state requires the full knowledge of the window function. In our context, such a window will corresponds to a spectral filtering operation as in Sec.~\ref{pseudonext} or to the spectrum of a reference single photon as in Sec.~\ref{othertech}. In addition, the bilinear structure of the chronocyclic Wigner distribution leads to the interference between negative and positive frequencies. Indeed, there is a new oscillatory signal located orthogonally to the axis of two distinct signals owing to this bilinear structure. Therefore, the CW distribution can be difficult to understand, which justifies the utilization of the CPW distribution. \\

\section{Spectral characterization of a single photon with generalized HOM interferometry}\label{sectionthree}

\subsection{Measurement of the chronocyclic Wigner distribution}\label{measurementFBS}

\begin{figure*}
 \begin{center}
\includegraphics[width=0.8\textwidth]{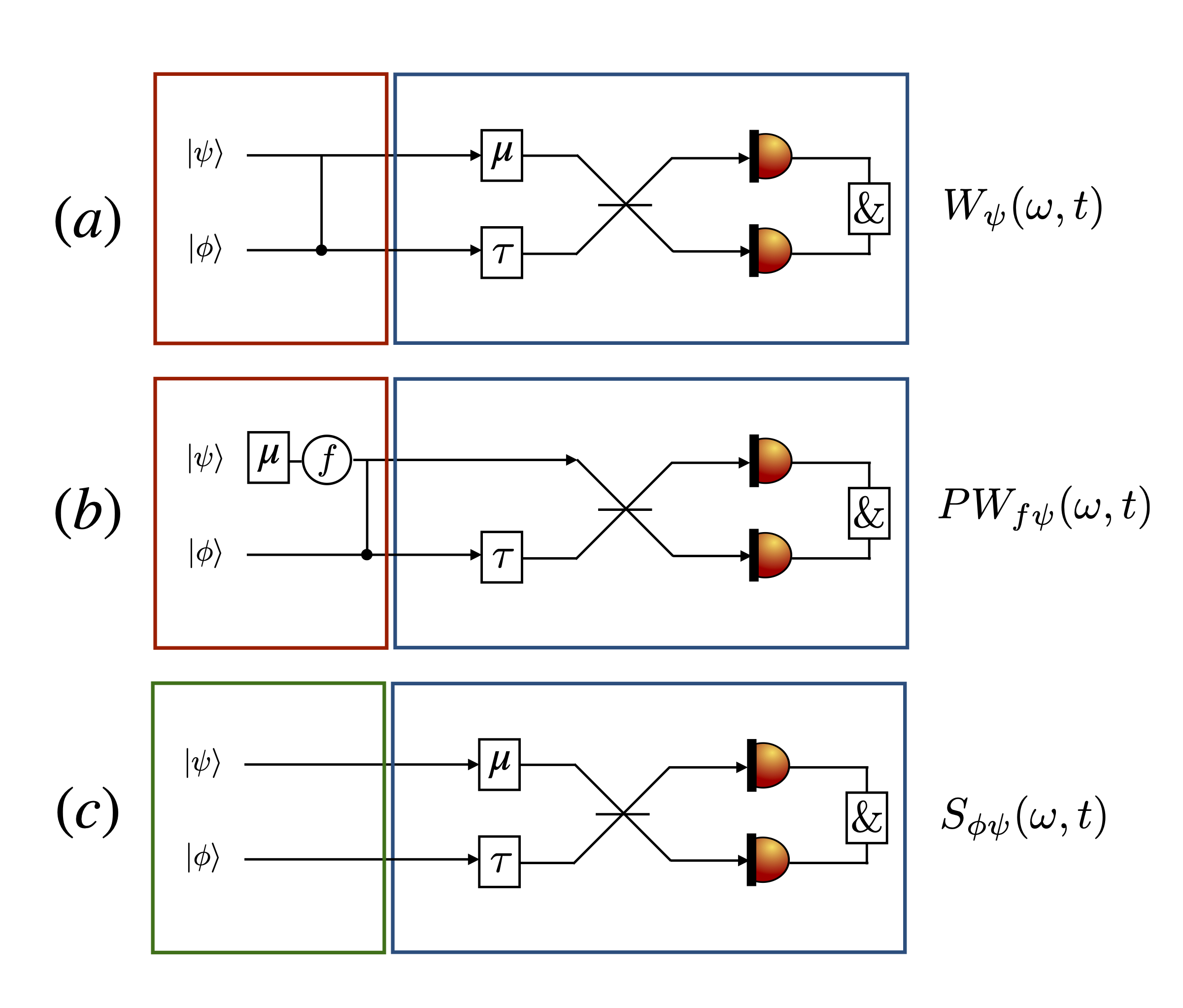}
\caption{\label{generalizedHOM} Schematic of the optical scheme for the spectral-temporal characterization of single photon states. (a) Two single photons are entangled with the frequency beam-splitter operation described by Eq.~(\ref{CNOTfrequgate})  (red rectangle). Then, time (resp. frequency) displacement operations are applied in spatial ports $a$ (upper) and $b$ (lower). The frequency entangled state is then recombined into a beam-splitter, and is then followed by coincidence measurement (in blue rectangle). (a) Measurement of the chronocyclic Wigner distribution. (b) Measurement of the pseudo-Wigner distribution, where spectral filtering is needed noted $f$ before the entangling operation. (c) Measurement of the spectrogram, where no frequency beam-splitter operation is necessary (enlightened by the green rectangle).}
\end{center}
\end{figure*}

We consider a single photon state $|\phi\rangle_{a}$ as a reference with spectrum $\phi$ in spatial port $a$ and the state to characterize $|\psi\rangle_{b}$ in spatial port $b$ (see Fig.~\ref{generalizedHOM}(a)). The full wave function of the initial separable state is:
\begin{equation}\label{initialstate}
|\Psi\rangle=|\phi\rangle_{a}\otimes |\psi\rangle_{b}= \int_{\mathbb{R}} \phi(\omega) d\omega |\omega\rangle_{a}\otimes \int_{\mathbb{R}} \psi(\omega') d\omega' |\omega'\rangle_{b}.
\end{equation}
The two single photons are now interacting with a frequency-balanced beam-splitter operation, which can be seen as a CNOT gate in this encoding. This operation is a non-linear interaction described by the following operator acting on the two single photon states:
\begin{equation}\label{CNOTfrequgate}
\hat{U}|\omega,\omega'\rangle_{ab}=|\frac{\omega+\omega'}{\sqrt{2}},\frac{\omega-\omega'}{\sqrt{2}}\rangle_{ab}.
\end{equation}
Using the Fourier transform relation between the time and the frequency degree of freedom, such a transformation can also be considered as a temporal beam-splitter:
\begin{equation}
\hat{U}|t,t'\rangle_{ab}=|\frac{t+t'}{\sqrt{2}},\frac{t-t'}{\sqrt{2}}\rangle_{ab}.
\end{equation}
The frequency beam-splitter operation manipulating here the time-frequency continuous variables can find some analogies with the quantum pulse gate manipulating time-frequency modes introduced in \cite{gil-lopez_universal_2021,ansari_tailoring_2018}.

\begin{figure}[h!]
 \begin{center}
\includegraphics[width=0.2\textwidth]{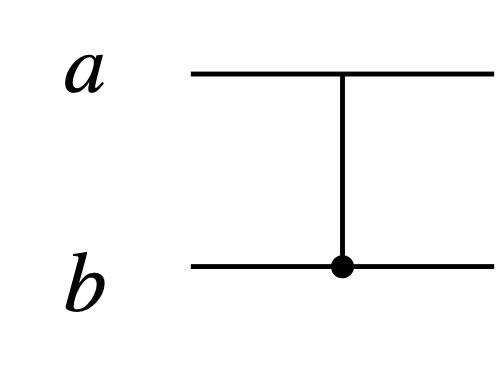}
\caption{\label{FrequencyBS} Quantum circuit diagram of the frequency beam-splitter operation. $a$ and $b$ denote different spatial paths. }
\end{center}
\end{figure}

This entangling gate is now written thanks to time-frequency operators, in order to make a direct mathematical connection with the widely known quadrature position-momentum formalism and also to identify the physical mechanisms that are needed for performing such an operation. The Hamiltonian corresponding to this gate is given by
\begin{multline}\label{balancedbs}
\hat{H}=\chi^{2}\hat{\Omega}_{a}\hat{\Omega}_{b}+\chi'^{2} \hat{T}_{a}\hat{T}_{b},\\
=(\frac{\chi}{\Delta\omega})^{2} \iint d\omega_{1}d\omega_{2}\omega_{1}\omega_{2}\hat{n}_{a}(\omega_{1})\hat{n}_{b}(\omega_{2})\\
+(\frac{\chi'}{\Delta t})^{2} \iint dt_{1}dt_{2}t_{1}t_{2}\hat{n}_{a}(t_{1})\hat{n}_{b}(t_{2}),
\end{multline}
where $\chi,\chi',\Delta\omega, \Delta t$ are constants whose explicit forms depend on the physical mechanism producing such a non-linear effect. To obtain a balanced-frequency beam-splitter, the relation between the coefficients must verify $(\frac{\chi}{\Delta\omega})^{2}=(\frac{\chi'}{\Delta t})^{2}=\theta=\pi/4$ which represents the angle characterizing the reflectivity and the transmission in the spectral-temporal domain $r=\text{cos}(\theta), t=\text{sin}(\theta)$. A quantum circuit diagram is represented in Fig.~\ref{FrequencyBS}. The mathematical structure of the frequency beam-splitter operation in Eq.~(\ref{balancedbs}) is closed to the one describing the Kerr interaction or a three-photon absorption effect. The introduction of time-frequency operators will help us to find the appropriate physical mechanism to perform such an operation.  \\

 The mathematical justification for the form of the Hamiltonian in Eq.~(\ref{balancedbs}) is as follows. The Hamiltonian of the beam-splitter operation is $\hat{H}=\theta(\hat{a}{b}^{\dagger}+\hat{a}^{\dagger}\hat{b})$, which can be expressed in terms of the quadrature position and momentum operators $\hat{a}=\hat{q}_{1}+i\hat{p}_{1}$ and $\hat{b}=\hat{q}_{2}+i\hat{p}_{2}$ as $\hat{H}=\theta(\hat{q}_{1}\hat{q}_{2}+\hat{p}_{1}\hat{p}_{2})$. For a balanced beam-splitter, the reflection and the transmission coefficients are equal to $r=t=1/\sqrt{2}$, and the angle is $\theta=\pi/4$. The beam-splitter acts on two quantum fields $|q_{1},q_{2}\rangle_{ab}$ in the same frequency mode but in two spatial modes $a$ and $b$  and is described by the following operator:
 \begin{equation}
 \hat{U}_{BS}|q_{1},q_{2}\rangle_{ab}=|\frac{q_{1}+q_{2}}{\sqrt{2}},\frac{q_{1}-q_{2}}{\sqrt{2}}\rangle_{ab}.
 \end{equation}
The formal mathematical analogy between quadrature and time-frequency continuous variables leads us to replace the quadrature operators with frequency and time operators. Setting aside this analogy in the mathematical structure, we must insist that the beam-splitter is a linear operation, while the frequency beam-splitter is not. The reader should be careful that in the denomination frequency beam-splitter has been employed in many contexts (see \cite{raymer_temporal_2020} for a review). For instance, we want to stress that the frequency beam-splitter is different from the frequency-dependent beam-splitter described in \cite{makarov_theory_2021,makarov_quantum_2021,joshi_frequency-domain_2020,kobayashi_frequency-domain_2016} or active beam-splitter \cite{raymer_interference_2010}.\\

The output state after the frequency beam-splitter can be written as
\begin{equation}
|\Psi\rangle=\iint d\omega d\omega' \phi(\omega)\psi(\omega') |\frac{\omega+\omega'}{\sqrt{2}},\frac{\omega-\omega'}{\sqrt{2}}\rangle_{ab}.
\end{equation}
After performing a change of variable, we obtain
\begin{equation}
|\Psi\rangle=\iint d\omega d\omega' \phi(\frac{\omega+\omega'}{\sqrt{2}})\psi(\frac{\omega-\omega'}{\sqrt{2}}) |\omega,\omega'\rangle_{ab},
\end{equation}
which is exactly the wave function of a photon pair produced by a SPDC process. Then, we retrieve a previous result found in \cite{fabre_generation_2020}, where we interpreted the SPDC process as two fictitious single photons that are entangled by a frequency beam-splitter operation modeling the non-linear Hamiltonian, without the explicit intervention of the classical pump. In such a process, $\phi,\psi $ denote the functions characterizing the energy conservation and the phase-matching condition depending on the collective variables $\omega_{\pm}=(\omega_{s}\pm\omega_{i})/\sqrt{2}$ of the signal and idler photons. From the observation that the phase-matching function of the SPDC process which depends on $\omega_{-}$ can be measured  with the generalized HOM interferometer, as shown in \cite{douce_direct_2013,boucher_toolbox_2015}, then it motivates and justifies the use of the frequency beam-splitter operation to measure the spectral function $\psi$ of the single photon of interest with such an interferometer. 

After temporal and frequency shifts are applied in the two spatial paths $a$ and $b$, the two single photons are then recombined into a balanced beam-splitter which is followed by coincidence measurement. The full optical scheme is represented in Fig.~\ref{generalizedHOM}(a).  The coincidence probability of two non-resolved temporal (or frequency) detectors, proven in Appendix \ref{appendixmixedstate} in the mixed state case, is found to be the CW distribution of the single photon of interest:
\begin{equation}
I(\tau,\mu)=\frac{1}{2}(1-\pi W_{\psi}(\tau,\mu)).
\end{equation}
In the expression of the coincidence probability, the spectral part depending on $\omega_{+}$ does not intervene: hence, the reference single photon state intervenes only to assist the two-photon interference effect. The CW distribution is reconstructed point by point, without the need for any post-calculation procedure. In Appendix \ref{appendixmixedstate}, we also establish that this protocol also allows  the reconstruction of the spectral distribution of single photon mixed states. The measurement of the CW at the single photon level has been performed with a compressive technique requiring a modal decomposition of the spectrum $\psi$ \cite{gil-lopez_universal_2021}. In our scheme, there is no need to spectrally shape the reference state. The optical difficulty resides in the frequency entanglement gate which is again a non-linear operation.

We now discuss a simple proposal for performing the frequency beam-splitter operation with four-wave mixing  and difference frequency generation. This operation is indeed non-energy preserving and needs to be performed auxiliary source of power. The photon pair generated by a SPDC process is filtered to obtain a separable state described by the wavefunction $|\psi\rangle =|\omega_{s},\omega_{i}\rangle$ where $\omega_{s}+\omega_{i}=\omega_{p}$ corresponds to the frequency of the pump. The signal photon along with a pump with frequency $\omega_{p}$ are the two inputs of a four-wave mixing process. The frequencies of the output photons are $\omega_{j}=2\omega_{s}-\omega_{p}=\omega_{s}-\omega_{i}$ and $\omega_{k}=2\omega_{p}-\omega_{s}=\omega_{s}+2\omega_{i}$ (see \cite{7021914}). The photon with frequency $\omega_{j}$ is one of the photon of interest. Then, a frequency difference generation process (as in \cite{walborn_quantum_2007}) between the photon $\omega_{k}$ and the photon from the SPDC process $\omega_{i}$ gives a photon with frequency $\omega_{s}+\omega_{i}$. In practice, after the four-wave mixing process, the seed (signal photon) is amplified and becomes a weak-coherent state, which decreases the visibility of the HOM coincidence. It is also the case for the stimulated parametric down conversion proposed in \cite{liscidini_stimulated_2013,https://doi.org/10.1002/lpor.201400057} for measuring the spectral function of photon pair. At this step, the state is under the form $|\omega_{s}+\omega_{i}\rangle |\omega_{s}-\omega_{i}\rangle$, but again the state $|\omega_{s}+\omega_{i}\rangle$ is not a single photon. The last needed operation is a frequency dilatation operation to obtain the state $|(\omega_{s}+\omega_{i})/\sqrt{2}\rangle |(\omega_{s}-\omega_{i})/\sqrt{2}\rangle$, which can be done with active beam-splitter \cite{raymer_interference_2010}. In addition, the DFG process with two-single photons is experimentally challenging. The deterministic realization of the frequency beam-splitter is difficult as it is the case for any operations manipulating degrees of freedom of single photons. The development of resonators and atomic systems can be beneficial for increasing the non-linearity at the single photon level.

Finally, in Appendix \ref{effectfinite}, we discuss that even if the frequency beam-splitter possesses a finite frequency bandwidth, it is again possible to reconstruct the chronocyclic Wigner distribution of interest. 

\subsection{Measurement of the chronocyclic pseudo-Wigner distribution}\label{pseudonext}
In this section, we present a method for measuring the chronocyclic pseudo-Wigner distribution of a single photon state. Starting from the single photon of interest $|\psi\rangle_{a}$, a frequency shift is applied, followed by a filtering operation described by the function $f$. The wave function, still written as $|\psi\rangle_{a}$, takes the following form:
\begin{equation}
|\psi\rangle_{a}=\int d\omega \psi(\omega+\mu) f(\omega) |\omega\rangle_{a}.
\end{equation}
Then, the protocol described in the previous section is applied and is fully represented in Fig.~\ref{generalizedHOM}(b). The input state of the beam-splitter is
\begin{equation}
|\Psi\rangle=\iint d\omega_{s}d\omega_{i} \psi(\omega_{-}+\mu)f(\omega_{-}) \phi(\omega_{+})e^{i\omega_{s}\tau} |\omega_{s},\omega_{i}\rangle_{ab}.
\end{equation}
Therefore, the coincidence probability measured with the generalized HOM interferometer is given by
\begin{multline}
I(\tau,\mu)=\frac{1}{2}(1-\int d\omega_{-} e^{2i\omega_{-} t} f^{*}(\omega_{-})f(-\omega_{-})
\psi(-\omega_{-}+\mu) \\
\cross\psi^{*}(\omega_{-}+\mu)),
\end{multline}
where we recognize the CPW distribution (see Eq.~(\ref{PseudoWigner})):
\begin{equation}
I(\tau,\mu)=\frac{1}{2}(1-\pi PW_{\psi}(\tau,\mu)).
\end{equation}
The ordering of the operations is important, the frequency shift must be applied before the filtering function because this last must not be shifted.  To obtain the chronocyclic Wigner distribution, post-calculations are required as indicated in Eq.~(\ref{recoverWigner}), but the CPW distribution can be a sufficient characterization of the single photon state. 

Note that the CPW distribution has been used in classical optics to erase the oscillation pattern which is a redundancy of information in this phase space representation. However, in the quantum case, we are interested to observe this oscillation pattern measured with generalized HOM interferometry because it indicates that the single photon state is pure. Indeed, the decrease in the purity of the quantum state can be detrimental to quantum computation applications. Apart from this purity consideration, the CPW distribution and other smooth versions could allow for the accurate identification  and easier identification  of unwanted side-band signals caused by faulty single photon gates. An electro optic modulator (EOM) or a pulse shaper could introduce new frequency components that can be observed by measuring the chronocyclic Wigner distribution. It is important to first erase the oscillation term coming from the quadratic form of the CW distribution,  to visualize in the CPW distribution potential side-bands caused by other physical processes. Thus, the CPW distribution is useful not only for error-diagnosis of classical fields but also of single photon state.

\subsection{Measurement of the spectrogram}\label{spectrogramsec}

In this section, we consider the case in which the frequency beam-splitter operation is not performed between the two single photons. Therefore, we consider two separable pure single photon states Eq.~(\ref{initialstate}) generated independently, or a frequency-separable photon pair produced by a SPDC process, entering in the generalized HOM interferometer. The protocol is represented in Fig.~\ref{generalizedHOM}(c). The expression of the coincidence probability in this case is
\begin{equation}
I(\tau,\mu)=\frac{1}{2}(1-S(\tau,\mu)),
\end{equation}
where $S$ corresponds to the spectrogram defined as:
\begin{equation}
S(\tau,\mu)=|\int_{\mathbb{R}} \phi(\omega_{s}-\mu)\psi(\omega_{s})e^{i\omega_{s}\tau} d\omega_{s}|^{2}.
\end{equation}
The quantity inside the absolute value is called the short-frequency Fourier transform, or the ambiguity function:
\begin{equation}\label{shortFourier}
X(\tau,\mu)=\int_{\mathbb{R}} \phi(\omega_{s}-\mu)\psi(\omega_{s})e^{i\omega_{s}\tau} d\omega_{s},
\end{equation}
which is related to the CW distribution using a double Fourier transform. In this last quantity, $\phi$ is a window function centered at frequency $\mu$ which cuts the spectrum $\psi$ around this frequency. Then, a subsequent Fourier transform is performed. The main difference between the two previous cases, is that the spectrum of the reference single photon state intervenes in the expression of the coincidence probability. The spectrogram is a well-known quantity in the time-frequency analysis of classical fields  \cite{dorrer_concepts_2005}. It can also be seen as the overlap of the CW distribution of the gate and the window; it is thus a blurred version of the CW distribution. The choice of the window function is crucial because it is not possible to achieve both time and frequency high resolution. Indeed, owing to the time-frequency inequality, time-energy one in the single photon regime \cite{fadel_time-energy_2021}, a wider frequency window improves the time domain resolution but decreases the frequency window. Interestingly, the spectrogram is a linear time-frequency representation \cite{467299} according to Eq.~(\ref{shortFourier}) and it is here obtained when no non-linear transformation is performed between the two single photons. The direct reconstruction of $\psi$ can be performed using the well-known phase retrieval technique \cite{dorrer_concepts_2005,stark}, as it has been done with the FROG technique.

 \begin{figure*}
 \begin{center}
\includegraphics[width=1\textwidth]{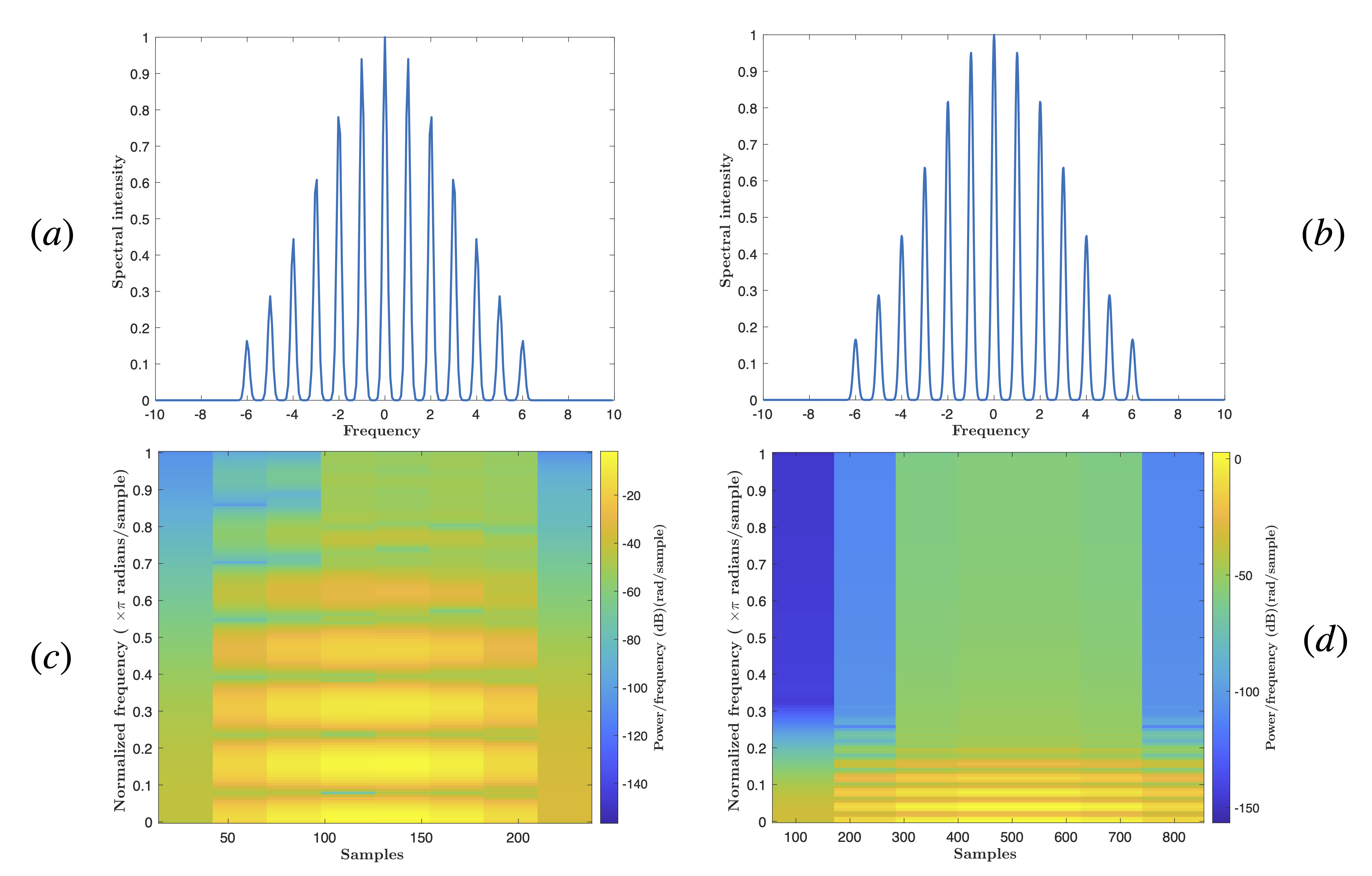}
\caption{\label{sampling} (a), (b)  Numerical simulation of the spectral intensity  with respect to the frequency (normalized with the frequency period) of frequency qudits defined by Eq.~(\ref{qudits}), $\kappa=0.1$ and its associated spectrogram (c), (d) with $N=256$ and $N=1024$. The central maximum frequency has been set to zero. The increase in the frequency resolution for $N=1024$ is at the cost of a decrease in the temporal resolution. As $N$ increases, the central normalized frequency of the qudit state decreases. In addition, there is no relevant information after the six lines representing the six positive frequencies of the qudit state. }
\end{center}
\end{figure*}

In the following, we will present the spectrogram of a frequency qudit state,  and how it can be used to measure the central frequency of each peak composing such a state. We consider the following qudit ($d=12$) state:
\begin{equation}\label{qudits}
|\psi\rangle=\sum_{n=-6}^{6} c_{n}|n\rangle,
\end{equation}
where $c_{n}=\text{exp}(-n^{2}\kappa^{2}/2)$, where $\kappa$ is the inverse of the width of the Gaussian distribution. The associated spectrogram of such frequency qudit state is represented for different resolutions ($N=256$) and ($N=1024$) in Fig.~\ref{sampling} (c) and (d). In these numerical simulations, the spectrum of the reference single photon $\phi$ is considered as a Hamming window (see Appendix \ref{spectroappendix} for more details) which could be engineered experimentally with a spatial light modulator (SLM) placed at the Fourier plane of a $4f$ optical scheme.  When the frequency resolution ($N$) is low, the frequency bandwidth of each peak is large, but the temporal bandwidth is lower compared to a higher sampling rate. Note that other single photon spectral windows can be used to measure the amplitude of each frequency peak with higher precision (compared to the Hamming window), such as the rectangular window, which can be useful for quantum protocols using an amplitude encoding.

In Appendix \ref{spectroappendix}, we study the case in which the single photon of interest state is a mixed one. The coincidence probability is not the spectrogram of the two-dimensional density matrix (see Eq.~(\ref{densitymatrix})) and we provide explanations for this fact. This result emphasizes the utility of the frequency beam-splitter gate which allows performing the tomography of both pure and mixed states.

\subsection{Other proposal for measuring the chronocyclic pseudo-Wigner distribution}\label{othertech}
In this section, we consider another possible frequency entanglement gate, which could be experimentally easier to implement and allows the measurement of the CPW distribution. The considered entanglement gate $\hat{C}_{x}$ is defined as follows:
\begin{equation}\label{Cxoperation}
\hat{C}_{x}|\omega\rangle_{a}|\omega'\rangle_{b}=|\omega\rangle_{a}|\omega'-\omega\rangle_{b}.
\end{equation}
This operation can be written with time and frequency operators as $\hat{C}_{x}=e^{i\hat{\omega}_{a}\otimes \hat{t}_{b}}$, where the corresponding Hamiltonian for this unitary transformation is
\begin{equation}
\hat{H}=\iint d\omega dt \ \omega t \hat{a}^{\dagger}(\omega)\hat{a}(\omega)\hat{b}^{\dagger}(t)\hat{b} (t). 
\end{equation}
The $\hat{C}_{x}$ gate can be seen as a frequency multimode cross-Kerr interaction between two spatial modes $a$ and $b$. However, note that there is a number operator in the frequency and time domains for the spatial mode $a$ (resp. $b$). After the frequency entanglement operation between the single photon of interest and the reference one (see Eq.~(\ref{initialstate})), the protocol is represented as well in Fig.~\ref{generalizedHOM}(a),  the coincidence probability measured using the generalized HOM interferometer is
\begin{multline}
I(\tau,\mu)=\frac{1}{2}(1-\text{Re}(\iint d\omega d\omega' \phi(\omega'+\mu)\psi(-\omega_{-}+\mu)\\
\cross \phi^{*}(\omega+\mu)\psi^{*}(\omega_{-}+\mu)e^{-2i\omega_{-}\tau})),
\end{multline}
with $\omega_{-}=\omega-\omega'$. We assume the most general form for the correlation function, and we decompose it thanks to a set of orthogonal functions $C^{j}$: $\phi(\omega_{+}+\mu-\omega_{-})\phi^{*}(\omega_{+}+\mu+\omega_{-})=\sum_{j}p_{j}C^{j}(\omega_{+}+\mu)C^{j}(\omega_{-})$, where $p_{j}\geq 0$ and $\sum_{j}p_{j}=1$ (an identical decomposition is given in Appendix \ref{effectfinite}, for more details) . If $\phi$ is a Gaussian function, then we have $p_{1}=1$ and the others $p_{i}$ are zero. The coincidence probability can be written as
\begin{multline}\label{equationmult}
I(\tau,\mu)=\frac{1}{2}(1-(\sum_{j}p_{j}\int_{\mathbb{R}} d\omega_{+} C^{j}_{1}(\omega_{+})\int_{\mathbb{R}} d\omega_{-} C^{j*}_{j}(\omega_{-})\\
\cross \psi(-\omega_{-}+\mu)\psi^{*}(\omega_{-}+\mu)e^{-2i\omega_{-}\tau})).
\end{multline}
We introduce $\alpha_{j}=\int d\omega_{+} C^{j}_{1}(\omega_{+})$ which is a known factor that depends only on the reference state.  We recognize in Eq.~(\ref{equationmult}) the sum of CPW distributions (see Eq.~(\ref{PseudoWigner})):
\begin{equation}
I(\tau,\mu)=\frac{1}{2}(1-\sum_{j}p_{j}\alpha_{j} PW_{C^{j}_{2}\psi}(\tau,\mu)).
\end{equation}
Note that, as in the measurement of the spectrogram, the spectral function of the reference here intervenes. Because $p_{j}\geq 0 $ and $\alpha_{j}$ are fully known, we can reconstruct the CPW distribution $PW_{C^{j}_{2}\psi}(\tau,\mu)$. This example shows the importance of the structure of the frequency entanglement for performing the tomography of the spectral-temporal single photon state, to reduce the number of post-processing calculations. Besides, no spectral filtering was used here in contrast to Sec.~\ref{pseudonext}, since the filtering function is the spectral distribution of the reference state and could be used as a spectral window to detect faulty optical components.

\section{Conclusion}\label{conclusion}
In this paper, we have explained how frequency entanglement, or not at all, between two single photons, frequency filtering and combined with  generalized HOM interferometry lead to the measurement of different chronocyclic phase-space distributions. We emphasize that in the presented cases, the more optical resources are used, through the complexity of the frequency entanglement gates, the less post-calculations are necessary to perform the full spectral-temporal tomography of single photons. In the case of the measurement of the spectrogram, the spectral distribution of the reference state is used as a spectral window  and this technique allows only measuring the spectral function of pure single photon states by using the retrieval phase algorithm.

A proposal to implement the frequency beam-splitter will be the subject of future work. Non-linear effects at the single photon level are experimentally challenging. The improvement of experimental platforms such as split-ring resonators \cite{joshi_frequency-domain_2020,lu_periodically_2019,yang_squeezed_2021,grassani_micrometer-scale_2015,kues_quantum_2019}, atoms coupled to two waveguides \cite{oehri_tunable_2015}, three-photon absorption effect (an example of such an effect in semiconductor materials is presented in \cite{benis_three-photon_2020}), are promising candidates towards the realization of such non-linear interactions. Indeed, the form of the Hamiltonian of the frequency beam-splitter suggests to use Kerr non-linear effect, which could be achieved with the mentioned experimental platforms. Furthermore, the recent development of EOMs \cite{karpinski_control_2021,chen_single-photon_2021} producing the frequency shift necessary for the characterization of single photon states, motivates the use of the generalized HOM interferometer. Besides, the chronocyclic pseudo-Wigner distribution has been used in a variety of applications involving classical fields, such as error diagnosis. In our context, the CPW distribution could be employed to detect faulty optical components such as EOM and pulse shaper, which are essential for quantum computation with time and frequency degrees of freedom as illustrated in \cite{lu_controlled-not_2019,lukens_frequency-encoded_2017,kues_quantum_2019}.

 As perspective, we could investigate the form of frequency entanglement gates and the spectral function of the reference state which would allow measuring each distribution belonging to the Cohen's class \cite{cohen_1989,467299}, other than the chronocyclic Wigner distribution and the spectrogram presented here. Besides, our proposal could also be applied to the transversal position-momentum continuous variables \cite{tasca_continuous_2011}.  We must emphasize the importance of such a frequency beam-splitter gate for quantum information protocols, such as for error correction \cite{fabre_generation_2020,fabre:tel-03191301}, in HOM metrology \cite{chen_hong-ou-mandel_2019,lyons_attosecond-resolution_2018,fabre_parameter_2021} or for universal time-frequency quantum computing which will be the subject of a future work.

\section*{Acknowledgements}

N.Fabre acknowledges useful discussion with Simone Felicetti, Marco Liscidini, Arne Keller and Perola Milman.
 N.Fabre acknowledges support from the project “Quantum Optical Technologies” carried out within the International Research Agendas programme of the Foundation for Polish Science, co-financed by the European Union under the European Regional Development Fund.

\bibliography{interactnlmsample}

\appendix
\section{Spectral characterization of mixed single photon state}\label{appendixmixedstate}

The density matrix of the bipartite system composed of the mixed single photon of interest with a pure reference state can be written as
\begin{equation}
\hat{\rho}=\iint  d\omega_{1} d\omega_{2} \rho(\omega_{1},\omega_{2}) |\omega_{1}\rangle\langle\omega_{2}| \otimes |\phi\rangle\langle \phi|.
\end{equation}
After the frequency beam-splitter operation (see Eq.~(\ref{CNOTfrequgate})), the new state also noted $\hat{\rho}$ is:
\begin{multline}
\hat{\rho}=\iint d\omega_{1} d\omega_{2} \phi(\frac{\omega_{1}+\omega}{\sqrt{2}}) \phi^{*}(\frac{\omega_{2}+\omega'}{\sqrt{2}}) 
 \rho(\frac{\omega_{1}-\omega}{\sqrt{2}},\frac{\omega_{2}-\omega'}{\sqrt{2}}) \\
 \cross |\omega_{1},\omega\rangle \langle\omega_{2},\omega'|, 
\end{multline}
The state is then introduced into the generalized HOM interferometer. After the frequency and temporal displacement operations, the state becomes:
\begin{multline}
\hat{\rho}=\iint d\omega_{1} d\omega_{2} \phi(\frac{\omega_{1}+\omega+\mu}{\sqrt{2}}) \phi^{*}(\frac{\omega_{2}+\omega'+\mu}{\sqrt{2}})\\ \times 
\rho(\frac{\omega_{1}-\omega-\mu}{\sqrt{2}},\frac{\omega_{2}-\omega'-\mu}{\sqrt{2}})e^{i\omega_{1}\tau}e^{-i\omega_{2}\tau}
|\omega_{1},\omega\rangle\langle\omega_{2},\omega'|. 
\end{multline}
The beam-splitter transforms the state $|\omega_{1},\omega\rangle\langle\omega_{2},\omega'|$ into:
\begin{multline}
\hat{a}^{\dagger}(\omega_{1})\hat{b}^{\dagger}(\omega)|0\rangle\langle0|\hat{a}(\omega_{2})\hat{b}(\omega')\rightarrow \frac{1}{2}(\hat{a}^{\dagger}(\omega_{1})+\hat{b}^{\dagger}(\omega_{1}))\\ \cross 
(\hat{a}^{\dagger}(\omega)-\hat{b}^{\dagger}(\omega))|0\rangle\langle0|(\hat{a}(\omega_{2})+\hat{b}(\omega_{2})) (\hat{a}(\omega')-\hat{b}(\omega')).
\end{multline}
The expression of the coincidence probability for non-resolved frequency measurement is $I(\tau,\mu)=\iint d\omega_{s}d\omega_{i} \langle\omega_{s},\omega_{i}|\hat{\rho}|\omega_{s},\omega_{i}\rangle$, and setting the frequency shift to $\mu\rightarrow \sqrt{2}\mu$ yields
\begin{multline}
I(\tau,\mu)=\frac{1}{4} \int_{\mathbb{R}} |\phi(\omega_{+}+\mu)|^{2} d\omega_{+} \int_{\mathbb{R}} d\omega_{-} 
 (\rho(\omega_{-}-\mu,\omega_{-}-\mu)\\
 +\rho(-\omega_{-}-\mu,-\omega_{-}-\mu)
 -\rho(\omega_{-}-\mu,-\omega_{-}-\mu)e^{2i\omega_{-}\tau}\\
 -\rho(-\omega_{-}-\mu,\omega_{-}-\mu)e^{2i\omega_{-}\tau}). 
\end{multline}
According to the normalization condition, $\int_{\mathbb{R}} \rho(\omega,\omega) d\omega=1$, we recognize the chronocyclic Wigner distribution of the mixed single photon state of interest:
\begin{equation}
I(\tau,\mu)=\frac{1}{2}(1-W_{\hat{\rho}}(\tau,\mu)).
\end{equation}
The same proof can be applied when a filtering function is applied before the frequency beam-splitter operation (see Sec.~\ref{pseudonext}), and shows in that case that the generalized HOM interferometer is a measurement of the CPW distribution of the spectral mixed single photon state.

\section{Effect of the finite bandwidth of the frequency beam-splitter gate on the spectral tomography of single photons}\label{effectfinite}
In this section, we investigate the effect of the finite bandwidth of the frequency beam-splitter operation, and it effect on single-photon tomography. The application of the CNOT gate on two single photon states is now described by the relation:
\begin{equation}
\hat{U}|\omega,\omega'\rangle_{ab}=U(\omega,\omega')|\omega_{+},\omega_{-}\rangle_{ab}.
\end{equation}
We investigate two possible forms of the function $U(\omega,\omega')$. The first is when $U$ is a separable function  $U(\omega,\omega')=U_{+}(\omega)U_{-}(\omega')$. The biphoton state after the CNOT operation is
\begin{equation}
|\psi\rangle=\iint d\omega d\omega' \phi(\omega_{+})\psi(\omega_{-}) U_{+}(\omega_{+})U_{-}(\omega_{-}) |\omega,\omega'\rangle_{ab}.
\end{equation}
The probability coincidence measured with the HOM interferometry is in that case:
\begin{equation}
I(\tau,\mu)=\frac{1}{2}(1-\frac{\beta}{\pi} W_{\psi.U_{-}}(\tau,\mu)).
\end{equation}
After calibration of the CNOT gate, namely the measurement of $U(\omega,\omega')$ and using a single photon source, the multiplicative factor $\beta=\int d\omega_{+} |\phi(\omega_{+})|^{2}  |U_{+}(\omega_{+})|^{2} $ can be determined. We obtain the chronocyclic Wigner distribution of the spectral function $\phi$ of interest multiplied by $U_{-}$. The reconstructing formula Eq.~(\ref{reconstruction}) allows obtaining the product $\psi(\omega)U_{-}(\omega)$. Assuming that $U_{-}$ is invertible, it is then possible to obtain $\psi(\omega)$.\\

In general, the function $U$ is not separable and can be decomposed into a set of orthogonal functions: $U(\omega,\omega')=\sum_{n,m} U_{nm}T_{n}(\omega)T_{m}(\omega')$, where $\int_{\mathbb{R}} T_{n}(\omega)T_{m}(\omega) d\omega= \delta_{nm}$, $\sum_{n}T_{n}(\omega)T_{n}(\omega')=\delta(\omega-\omega')$. After performing a singular value decomposition on the matrix $U_{nm}=(ZpV)_{nm}$, where $p$ is a diagonal matrix and $Z,V$ are unitaries one, we find:
\begin{equation}
U(\omega,\omega')=\sum_{j} p_{j} p_{1}^{j}(\omega)p_{2}^{j}(\omega'),
\end{equation}
where $\sum_{j}p_{j}=1$, $p^{j}_{1}(\omega)=\sum_{n}Z_{nj}T_{n}(\omega)$, $p^{j}_{2}(\omega)=\sum_{m}V_{km}T_{m}(\omega')$. The case where $p_{1}=1$ and the others are zero is the previous separable case. We assume that the function $C$ is fully known. The expression of the coincidence probability is expressed as follows:
\begin{equation}
I(\tau,\mu)=\frac{1}{2}(1-\sum_{j,k}p_{j}p_{k} \beta_{j,k} \overline{W}_{g_{j}g_{k}}(\tau,\mu)),
\end{equation}
where we have defined $\beta_{j,k}=\int_{\mathbb{R}} d\omega |\phi(\omega)|^{2}p_{1}^{j}(\omega)p_{1}^{k*}(\omega)$, $g_{j}(\omega)=p_{2}^{j}(\omega)\psi(\omega)$ and the cross chronocyclic Wigner distribution:
\begin{equation}\label{biche}
\overline{W}_{g_{j}g_{k}}(\tau,\mu)=\int_{\mathbb{R}} e^{2i\omega'\tau} g_{j}(\omega-\omega')g^{*}_{k}(\omega+\omega') d\omega'.
\end{equation}
We now give the main idea behind the reconstruction of the correlation spectral function of the single photon of interest: $C_{\psi}(\omega,\omega')=\psi(\omega-\omega')\psi^{*}(\omega+\omega')$. Once the coincidence probability is measured (see Eq.~(\ref{biche})), we start by inverting the relation $Y=X^{\top}AX$, where $Y_{jk}=\sum_{j,k}p_{j}p_{k} \beta_{j,k} \overline{W}_{g_{j}g_{k}}(\tau,\mu)$, the column vector $X_{j}=p_{j}$ and the matrix $A_{jk}= \beta_{j,k} \overline{W}_{g_{j}g_{k}}(\tau,\mu)$. Then, we obtain the matrix $A$. Because the coefficients $ \beta_{j,k}$ are known, because we suppose that the reference and the quantum material producing the frequency beam-splitter are fully characterized, we access to the cross-Wigner distribution $ \overline{W}_{g_{j}g_{k}}(\tau,\mu)$. A Fourier numerical transform of this last quantity allows the correlation function to be accessed:
\begin{multline}
C_{jk}(\omega,\omega')=g_{j}(\omega-\omega')g_{k}^{*}(\omega+\omega')
\\=p_{j}(\omega-\omega')p_{k}^{*}(\omega+\omega')C_{\psi}(\omega,\omega').
\end{multline}
For $p_{j}(\omega-\omega')p_{k}^{*}(\omega+\omega')\neq 0$, we obtain the correlation function $C_{\psi}(\omega,\omega')$.

To conclude, if the frequency bandwidth of the reference and the single photon of interest are much lower than those of the frequency beam-splitter, we can consider the case study in Sec.~\ref{measurementFBS}. Otherwise, the reconstruction of the full tomography of the spectral single photon state is also possible, but requires more post-processing calculations.\\

\section{Additional information about the spectrogram}\label{spectroappendix}
In this appendix, we provide additional information about the measurement of the spectrogram at the single photon level.

The spectral function of the reference single photon to perform the spectrogram which is used as a Hamming window is given by
\begin{equation}
\phi(\omega)=\left\{
    \begin{array}{ll}
    \text{cos}^{2}(\frac{\pi \omega }{\Delta\omega}) & \text{if} \ |\omega| \leq \Delta\omega/2 \\
    0 & \text{if} \ |\omega|\geq \Delta\omega/2 .
    \end{array}
\right.
\end{equation}
In practice, such a function is actually discretized,
\begin{equation}
\phi[n]= \phi[\frac{\Delta\omega}{N}(n-\frac{N}{2})]= \text{sin}^{2}(\frac{\pi n}{N}). 
\end{equation}
This spectral function could be engineered with the association of EOM and pulse shaping operations \cite{lukens_frequency-encoded_2017}, or by mapping the frequency to the spatial variable with a grating followed by a SLM placed at the Fourier plane of a $4f$ optical scheme.

We now provide the case where the single photon of interest is mixed, the reference being pure. The coincidence probability yields
\begin{multline}
I(\tau,\mu)=\frac{1}{2}(1-\iint d\omega_{s}d\omega_{i} \rho(\omega_{s}-\mu,\omega_{i}-\mu)\phi(\omega_{i})\phi^{*}(\omega_{s})\\
\times e^{i(\omega_{s}-\omega_{i})\tau}).
\end{multline}
This function can be seen as the diagonal element of the four-dimensional spectrogram $S(\mu,\mu,\tau,\tau)$ of a two-dimensional function $\rho(\omega_{s},\omega_{i})$ with the window function $\phi(\omega_{s})$.
Nevertheless, two parameters are lacking to reconstruct the density matrix $\rho(\omega_{s},\omega_{i})$  with the phase-retrieval algorithm.

\end{document}